\documentclass{PoS}
\usepackage{amsmath,amssymb,graphicx}

\newcommand{\partialdisplay}[1]{\frac{\partial}{\partial #1}}
\newcommand{\D}{\rho}

\title{Two-loop $D$-dimensional unitarity and dual conformal symmetry}

\ShortTitle{$D$-dimensional dual conformal symmetry}

\author{Zvi Bern$^{\, a}$, Michael Enciso$^{\, a}$, Harald Ita$^{\, b}$, \speaker{Mao Zeng}$^{\, a}$\\
        $^a$Mani L. Bhaumik Institute for Theoretical Physics, Department of Physics and Astronomy, University of California at Los Angeles, California 90095\\
        $^b$ Physikalisches Institut, Albert-Ludwigs-Universit\"at, Freiburg\\
D-79104 Freibug, Germany\\
E-mail: \email{bern@physics.ucla.edu}, \email{menciso@physics.ucla.edu}, \email{harald.ita@physik.uni-freiburg.de}, \email{zengmao@physics.ucla.edu}}

\abstract{In this talk we show that dual conformal symmetry has unexpected applications to Feynman integrals in dimensional regularization. Outside $4$ dimensions, the symmetry is anomalous, but still preserves the unitarity cut surfaces. This generally leads to differential equations whose RHS is proportional to $(d-4)$ and has no doubled propagators. The stabilizer subgroup of the conformal group leads to integration-by-parts (IBP) relations without doubled propagators. The above picture also suggested hints that led us to find a nonplanar analog of dual conformal symmetry.}

\FullConference{Loops and Legs in Quantum Field Theory (LL2018)\\
		29 April 2018 - 04 May 2018\\
		St. Goar, Germany}

\begin{document}

\section{Introduction}
Dual conformal symmetry is a hidden symmetry of planar $\mathcal N=4$ super-Yang-Mills (sYM) amplitudes, as well as many individual Feynman integrals \cite{0807.1095, hep-th/0607160, 1004.5381, 1103.1016, 1105.2011, 1404.2922, Broadhurst:1993ib}. The symmetry is exact only in integer dimensions, typically $4$ dimensions, but the anomalies in general $d$ dimensions are precisely understood in the case of $\mathcal N=4$ sYM amplitudes \cite{0807.1095}, which strongly constrains the infrared structure of the amplitudes in dimensional regularization. Here we will show that the anomalies for individual integrals also have a rather simple structures, and connect generalized unitarity with integration-by-parts (IBP) reduction and differential equations (DEs), both of which are important tools for evaluating multi-loop integrals. This talk is mainly based on Ref.~\cite{Bern:2017gdk}.

In Section \ref{sec:ibp-de}, we introduce recent new approaches to IBP reduction and DEs, which avoid doubled propagators. The key challenges in these approaches, related to the structures of unitarity cut surfaces, are discussed. In Section \ref{sec:dual}, we present unexpected simplifications due to insights from $\mathcal N=4$ sYM theory, in particular dual conformal symmetry and momentum twistors. Examples at one and two loops are given. In Section \ref{sec:nonplanar}, we further exploit the connections between dual conformal symmetry and unitarity cut surfaces, and identify an analog of dual conformal symmetry for nonplanar Feynman integrals.

\section{IBP relations and differential equations without doubled propagators}
\label{sec:ibp-de}
In dimensional regularization, total derivatives integrate to zero,
\begin{equation}
0 = \int d^d l \, \partialdisplay{l^\mu} 
  \frac{v^\mu \, \mathcal N}{\prod_j \D_j}\,, 
\label{eq:ibp}
\end{equation}
where $\D_j$ are propagator denominators, $\mathcal N$ is an arbitrary numerator, and $v^\mu$ is a Lorentz-vector with polynomial dependence on internal and external momenta. For illustration, the above equation is written down in the one-loop case, but it is trivial to generalize to the multi-loop case.

Explicitly evaluating the total derivative in the integrand of Eq.~\eqref{eq:ibp} gives us integration-by-parts (IBP) relations \cite{Chetyrkin:1981qh}, which are linear relations between Feynman integrals. Solving the linear system \cite{hep-ph/0102033}, all integrals with a given propagator structure are usually reduced to a small number of master integrals. This is a ubiquitous step in many multi-loop calculations.

However, there are clear redundancies in this procedure. Because of derivatives acting on propagator denominators in Eq.~\eqref{eq:ibp}, the IBP relations contain ``auxiliary integrals'' with doubled (i.e.\ squared) propagators, while most integrals that actually arise from Feynman diagrams do not have doubled propagators.\footnote{The exceptions are diagrams with internal self-energy insertions, but they affect only a small subset of diagram topologies.}
To address the problem, Gluza, Kajda, and Kosower (GKK) \cite{Gluza:2010ws} proposed the following extra condition on $v^\mu$ in Eq.~\eqref{eq:ibp},
\begin{equation}
v^\mu \partialdisplay{l^\mu} \rho_j = f_j \rho_j,
\label{eq:GKK}
\end{equation}
with $f_j$ being a polynomial, for every inverse propagator $j$ (no summation). This cancels doubled propagators and gives IBP relations between integrals with single propagators.

A simple but important geometric interpretation of Eq.~\eqref{eq:GKK} is pointed out by Ref.~\cite{Ita:2015tya}: $v^\mu$ is a polynomial tangent vector of the unitarity cut surface defined by $\rho_j = 0$.\footnote{There are different unitarity cuts, depending on which inverse propagators are set to zero. But it is easy to see that $v^\mu$ is tangent to all unitarity cut surfaces.} This observation allowed the complete solution of Eq.~\eqref{eq:GKK} for all one-loop integral topologies as well as a subset of planar two-loop integral topologies \cite{Ita:2015tya}. Beyond these simplest cases,
progress has been made using computational algebraic geometry and linear algebra \cite{Gluza:2010ws, 1111.4220, Larsen:2015ped, Zhang:2016kfo, Abreu:2017xsl, Abreu:2017hqn, Boehm:2018fpv}. But it is desirable to find analytic solutions, which will be the topic of Section \ref{sec:dual}.

A related problem, for the purpose of evaluating master integrals, is the construction of differential equations in dimensional regularization \cite{Kotikov:1990kg, Bern:1993kr, Remiddi:1997ny}. Here we compute derivatives of the master integrals w.r.t.\ external momenta $p_i^\mu$,
\begin{equation}
\beta^\mu \partialdisplay{p_i^\mu} \int d^d l \frac{\mathcal N}{\prod_j \rho_j},
\label{eq:de}
\end{equation}
and the resulting integrals are again reduced to the master integrals using IBP. To avoid doubled propagators generated by differentiation, Refs.~\cite{1702.02355, 1712.03760} modified Eq.~\eqref{eq:de} by adding IBP relations that integrate to zero,
\begin{equation}
\int d^d l \left[ \beta_i^\mu \partialdisplay{p_i^\mu} \frac{\mathcal N}{\prod_j \rho_j} + \partialdisplay{l^\mu} \frac{v^\mu \, \mathcal N}{\prod_j \rho_j} \right],
\label{eq:deDressed}
\end{equation}
and imposing a generalization of the GKK condition Eq.~\eqref{eq:GKK},
\begin{equation}
\left( \beta_i^\mu \partialdisplay{p_i^\mu} + v^\mu \partialdisplay{l^\mu} \right) \rho_j = f_j \, \rho_j \, ,
\label{eq:GKK-de}
\end{equation}
This again has a geometric interpretation: the terms in the bracket on the LHS of Eq.~\eqref{eq:GKK-de} is a tangent vector of unitarity cut surfaces in the space of both internal and external momenta. For given $\beta_i^\mu$, finding the ``compensating'' $v^\mu$ that solves Eq.~\eqref{eq:GKK-de} is again a problem that can be solved by computational algebraic geometry, but as we will see, interesting analytic solutions arise from dual conformal transformation.

\section{Analytic insights from $\mathcal N=4$ super-Yang-Mills theory}
\label{sec:dual}
\subsection{Dual conformal symmetry and differential equations}
Inspired by the Wilson loop-amplitudes duality \cite{0807.1095} in $\mathcal N=4$ sYM, a planar Feynman diagram can be mapped to dual space, where each momentum line becomes a difference between two dual coordinates.
\begin{figure}
  \centering
  \includegraphics[width=0.3\textwidth]{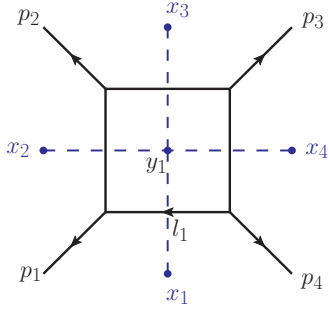}
  \caption{The one-loop box diagram with outgoing external momenta $p_1, \, p_2, \, p_3, \, p_4$. We introduce dual coordinates $x_1, x_2, x_3, x_4, y_1$.}
  \label{fig:box-dual}
\end{figure}
An example is the one-loop box diagram in Fig.~\ref{fig:box-dual}. The external momenta are mapped to differences between dual coordinates, as
\begin{equation}
p_1^\mu = x_2^\mu - x_1^\mu, \quad p_2^\mu = x_3^\mu - x_2^\mu, \quad p_3^\mu = x_4^\mu - x_3^\mu, \quad p_4^\mu = x_1^\mu - x_4^\mu \, .
\label{eq:boxMap1}
\end{equation}
Similarly, the $4$ internal lines are also mapped to difference between dual coordinates,
\begin{equation}
l_1^\mu = y_1^\mu - x_1^\mu, \quad l_1^\mu-p_1^\mu = y_1^\mu  - x_2^\mu, \quad l_1^\mu - p_1^\mu - p_2^\mu = y_1^\mu - x_3^\mu, \quad l_1^\mu + p_4^\mu = y_1^\mu - x_4^\mu \, .
\label{eq:boxMap2}
\end{equation}
Eq.~\eqref{eq:boxMap1} guarantees the overall momentum conservation. Momentum conservation at each vertex is also manifest from Eqs~\eqref{eq:boxMap1} and \eqref{eq:boxMap2}. Unitarity cuts now have a clear interpretation in dual coordinate space, as follows: \emph{cutting a propagator, i.e.\ setting a propagator on-shell, is equivalent to setting two dual coordinates to be light-like separated.}
 
For diagrams that appear in $\mathcal N=4$ sYM, the loop integrand is invariant under conformal transformation of the dual coordinates.\footnote{The integrand needs to be in an appropriate representation that makes the symmetry manifest.} This is known as dual conformal symmetry, and is distinct from the ordinary conformal symmetry of the theory.
The \emph{momentum-twistor formalism} \cite{Hodges:2009hk} makes dual conformal symmetry manifest. In this formalism, every dual coordinate corresponds to a line in momentum twistor space, and two dual coordinates are light-like separated if and only if they correspond to two lines that intersect each other in momentum twistor space. This gives a manifestly dual conformal invariant picture of unitarity cuts in 4 dimensions! The implication for our study is that infinitesimal dual conformal transformations always generate tangent vectors to unitarity cut surfaces, and give solutions to Eq.~\eqref{eq:GKK-de}.\footnote{In fact, the tangent vectors have polynomial components, as is clear from the explicit forms of the generators, e.g.\ in Eq.~\eqref{eq:boostFormula}}

However, the momentum twistor formalism is specific to $4$ dimensions. Fortunately, the argument carries over to $d$ dimensions. Consider a conformal boost in $d$ dimensions with parameter $b^\mu$,
\begin{equation}
\Delta x^\mu = \frac 1 2 x^2 b^\mu - (b \cdot x) x^\mu \, .
\label{eq:boostFormula}
\end{equation}
The squared separation between two dual coordinates transforms as
\begin{equation}
\Delta (x_1 - x_2)^2 = -b \cdot (x_1 + x_2) (x_1 - x_2)^2 \, .
\label{eq:propWeightBoost}
\end{equation}
Identifying $(x_1-x_2)^2$ with $\rho_j$ in Eq.~\eqref{eq:GKK-de}, the transformation generates a solution to Eq.~\eqref{eq:GKK-de} with $f_j = -b \cdot (x_1 + x_2)$. On the unitarity cut surface with $\rho_j (x_1-x_2)^2 = 0$, the RHS of Eq.~\eqref{eq:propWeightBoost} is zero, so the light-like separation between dual points is maintained. This is not surprising, as conformal transformations preserve the causal structure of spacetime.

The box diagram actually appears with the numerator
\begin{equation}
s t = (p_1+p_2)^2 (p_2+p_3)^2 = (x_3-x_1)^2 (x_4-x_2)^2
\end{equation} in the color-ordered amplitudes in $\mathcal N=4$ sYM. Changing the integration variables from the loop momentum $l^\mu$ to the dual coordinate $y_1^\mu$, the box integral is written as
\begin{equation}
I^{\rm box} = \int d^d y_1 \, \frac{(x_3-x_1)^2 (x_4-x_2)^2} {(y_1-x_1)^2 (y_1-x_2)^2 (y_1-x_3)^2 (y_1-x_4)^2} \, .
\end{equation}
The numerators and denominators in the integrand all transform with a local weight given in Eq.~\eqref{eq:propWeightBoost} under a conformal boost in the dual coordinate space. The integration measure varies as
\begin{equation}
\Delta \left( d^d y_1 \right) = -d (b \cdot y_1) d^d y_1 \, .
\end{equation}
Adding up all the weights, the box integral transforms as
\begin{equation}
\Delta I^{\rm box} = \int d^d y_1 \, (4-d) (b \cdot y_1) \frac{(x_3-x_1)^2 (x_4-x_2)^2} {(y_1-x_1)^2 (y_1-x_2)^2 (y_1-x_3)^2 (y_1-x_4)^2} \, .
\label{eq:DeltaBox}
\end{equation}
If we formally set $d=4$, ignoring the need to use dimensional regularization to regulate infrared divergences of the integral, then Eq.~\eqref{eq:DeltaBox} is exactly the statement that the box integral is dual conformal invariant. Keeping full dependence in $d$ and choose a conformal transformation generator that varies $t$ but leaves $s$ invariant, we derive the following differential equation
\begin{equation}
2(s+t) t \partialdisplay{t} \left( s t I^{\rm box} \right) = \epsilon \left[ -2s \left( s t I^{\rm box} \right) + 4st I^{\rm tri,t} - 4st I^{\rm tri,s} \right],
\end{equation}
where the RHS is proportional to $\epsilon = (4-d)/2$, and contains a $t$-channel triangle and an $s$-channel triangle. This is exactly Henn's $\epsilon$-factorized form of the differential equations \cite{1304.1806}, which has proven to be a powerful tool to obtain analytic $\epsilon$-expansion of Feynman integrals. The standard derivation of the above DEs involve IBP reduction of integrals with doubled propagators, which is entirely side-stepped in our method.

The above argument generalizes to any integral that has dual conformal invariance in 4 dimensions, and we always obtain DEs whose RHS is proportional to $\epsilon$ and free of integrals with doubled propagators. This gives a precise connection between the symmetry properties of the integrals and analytic properties of the $\epsilon$ expansions of the integrals.

\subsection{Stabilizer subgroup and integration by parts}
In this section, we will not restrict ourselves to integrals that are dual conformal in $4$ dimensions. As a result, the anomalies under dual conformal transformations will no longer be proportional to $(d-4)$, but still will be given by integrals without doubled propagators. To generate IBP relations instead of DEs, using $v^\mu$ satisfying Eq.~\eqref{eq:GKK} instead of Eq.~\eqref{eq:GKK-de}, we need dual conformal transformations which only act non-trivially on loop momenta but leaves external momenta unchanged. In terms of dual coordinates, we need the stabilizer subgroup of the conformal group which leaves the external dual coordinates, e.g.\ $x_i^\mu$ in Fig.~\ref{fig:box-dual}, unchanged.

\begin{figure}
  \centering
  \includegraphics[width=0.3\textwidth]{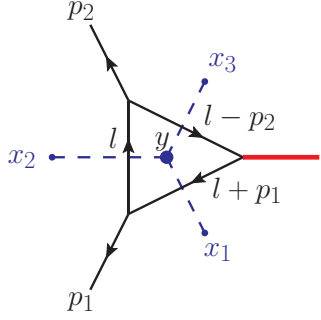}
  \caption{The one-loop triangle with outgoing external momenta $p_1, \, p_2, \,-p_1-p_2$ and dual points $x_1,x_2,x_3$. The only massive line is the external leg $(p_1+p_2)^2=s$.}
  \label{fig:triangle}
\end{figure}

A simple example is the one-loop triangle integral, shown in Fig.~\ref{fig:triangle} together with the mapping to dual coordinate space.
We fix the translation invariance of dual coordinate space by fixing $x_2$ as the origin, so the dual coordinates are given by explicit expressions,
\begin{equation}
x_1^\mu = -p_1^\mu, \qquad x_2^\mu = 0, \qquad x_3^\mu = p_2^\mu, \qquad y^\mu = l^\mu \, .
\end{equation}
Now we try to write down a conformal symmetry generator which keeps the external dual coordinates, $x_1^\mu, x_2^\mu, x_3^\mu$ unchanged. We consider the sum of a conformal boost Eq.~\eqref{eq:boostFormula} and a scaling transformation
\begin{equation}
\Delta x^\mu = \beta x^\mu \, .
\end{equation}
It can be checked that with the choice
\begin{equation}
\beta = s = (p_1 + p_2)^2, \qquad b^\mu = -2 (x_1^\mu + x_3^\mu) = -2 (p_2^\mu - p_1^\mu),
\end{equation}
all external dual coordinates are invariant, while the internal dual coordinate $y$ transforms as
\begin{equation}
\Delta y^\mu = \Delta l^\mu = -l^2 (x_1^\mu + x_3^\mu) + \left[s + 2l \cdot (x_1+x_3) \right] l^\mu \, .
\end{equation}
Identifying this as $v^\mu$ in Eq.~\eqref{eq:GKK}, the total divergence Eq.~\eqref{eq:ibp} evaluates to the following IBP relation involving the scalar triangle integral $I^{\rm tri}$ and the s-channel bubble integral $I^{\rm bub}_{(s)}$,
\begin{equation}
(d-4) s \, I^{\rm tri} + 2(d-3) I^{\rm bub}_{(s)},
\end{equation}
which is easily verified by explicit evaluation of the triangle and bubble integrals. The IBP relation is obtained in a clean way, as auxiliary integrals with doubled propagators are avoided.

\begin{figure}
  \centering
  \includegraphics[width=0.38\textwidth]{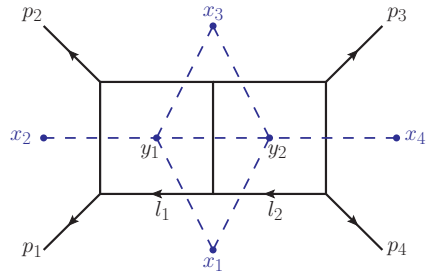}
  \caption{The massless double box integral, and the mapping to dual coordinates.}
  \label{fig:dbox-dual}
\end{figure}

The above method is only sensitive to the external dual coordinates but is agnostic about the loop order, and directly applies to e.g.\ nontrivial two-loop integrals. An example is the double box integral in Fig.~\ref{fig:dbox-dual}.
Again we identify conformal transformations that leave the external dual coordinates $x_i^\mu$ invariant. Our results, in a few lines \cite{Bern:2017gdk}, reproduces nearly one page of expressions in Ref.~\cite{Gluza:2010ws}, and reduce all tensor integrals to two master integrals modulo sub-topology integrals. More complicated two-loop integrals, such as the pentabox, can also be treated with generalizations of the above method \cite{Bern:2017gdk}.

\section{Nonplanar analog of dual conformal symmetry}
\label{sec:nonplanar}
For nonplanar diagrams, there is no obvious mapping to dual coordinates. However, important analytic properties of planar $\mathcal N=4$ sYM loop integrands, such as having only logarithmic singularity and no poles at infinity, surprisingly carries over to the nonplanar sector of the theory \cite{1410.0354, 1412.8584, 1512.08591}. So it is natural to speculate that there may be a nonplanar analog of dual conformal symmetry as well.

In previous sections, we connected the following two concepts, (i) dual conformal symmetry, and (ii) polynomial tangent vectors of unitarity cut surfaces. Point (ii) makes no direct reference to planarity, as unitarity cut surfaces are defined for any loop integral. So our strategy to attack the problem is as follows: \emph{find a polynomial tangent vector of unitarity cut surfaces of nonplanar diagrams, and check whether it generates a hidden symmetry of the integral.}
\begin{figure}
  \centering
  \includegraphics[width=0.38\textwidth]{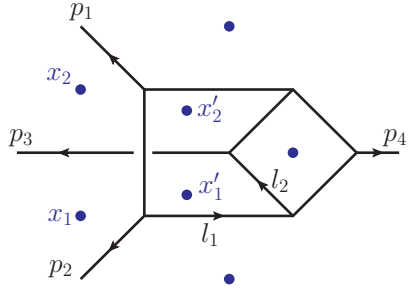}
  \caption{The nonplanar double box integral.}
  \label{fig:npdbox}
\end{figure}
An example is the nonplanar double box integral in Fig.~\ref{fig:npdbox}. We attempt to introduce dual coordinates just as in the planar case, indicated by blue dots in the figure. A strange new feature appears: the external leg $p_3^\mu$ is written as the difference between dual coordinates in two different ways,
\begin{equation}
p_3^\mu = x_2^\mu - x_1^\mu = x_2^{\prime \, \mu} - x_1^{\prime \, \mu} \, .
\label{eq:consistency}
\end{equation}
An infinitesimal conformal boost of the dual coordinates $x_1^\mu, x_2^\mu, x_1^{\prime \, \mu}, x_1^{\prime \, \mu}$ no longer gives a meaningful transformations of the external momentum $p_3^\mu$, unless the second equality in Eq.~\eqref{eq:consistency} is maintained. This rules out almost all the conformal boost generators given in Eq.~\eqref{eq:boostFormula}, \emph{except} the one with parameter $b^\mu = p_3^\mu$,
which preserves Eq.~\eqref{eq:consistency}, in fact with $\Delta p_3^\mu = 0$, following a simple calculation starting from Eq~\eqref{eq:propWeightBoost}.
We thus obtain a consistent transformation of external and internal momenta while preserving the unitarity cut surfaces. Remarkably, with the numerator $(p_1 + p_2)^2 (p_1+p_3)^2 (l_1-p_3)^2$, the nonplanar double box integral in $4$ dimensions is formally invariant under this transformation.\footnote{The above numerator is exactly the one found in Ref.~\cite{1512.08591} which manifests the simple analytic properties of the nonplanar $\mathcal N=4$ sYM loop integrand.} We have identified a novel hidden symmetry of a nonplanar integral that appears in the two-loop amplitudes of $\mathcal N=4$ sYM at finite $N_c$. Since this talk was given, the symmetry has been extended to integrals in nonplanar $\mathcal N=4$ sYM amplitudes at 2 loops and 5 points \cite{Bern:2018oao}, and implications at the integrated level was investigated by Ref.~\cite{Chicherin:2018wes}.

\section{Conclusions}
IBP reduction is a major computational bottleneck in simplifying complicated loop amplitudes relevant for collider physics, and is an essential step in the differential equation method for computing master integrals. Novel methods based on generalized unitarity and computational algebraic geometry have shown great promise. Additional analytic understanding is desirable from a theoretical points of view, and compact analytic results are also beneficial for practical calculations. The unexpected simplification from dual conformal symmetry exactly provides such analytic input, and simplifies IBP reduction as shown in one- and two-loop examples. On a related front, dual conformal transformations in $d$ dimensions give a simplified construction of differential equations, and provides an appealing new perspective on Henn's $\epsilon$ form of differential equations \cite{1304.1806}, originally motivated by the polylogarithm structures of many loop integrals.

By exploring the connection between dual conformal transformations and the tangent vectors of unitarity cut surfaces, we also uncovered a nonplanar analog of dual conformal symmetry. Besides the practical utility of constructing IBPs and DEs for nonplanar integrals in $d$ dimensions, the new symmetry also suggests possible nonplanar generalizations of the representation of $\mathcal N=4$ sYM integrands related to on-shell diagrams, the positive Grassmannian \cite{1212.5605} and the amplituhedron \cite{1312.2007, 1312.7878}.

\end{document}